\documentclass[sigconf]{acmart}
\usepackage{amsthm,color}
\usepackage{multirow, tabularx}
\usepackage{graphicx}
\usepackage{placeins}
\usepackage{subfig}
\theoremstyle{definition}

\newtheorem{definition}{Definition}

\AtBeginDocument{%
  \providecommand\BibTeX{{%
    \normalfont B\kern-0.5em{\scshape i\kern-0.25em b}\kern-0.8em\TeX}}}

\setcopyright{acmcopyright}
\copyrightyear{2020}
\acmYear{2020}
\acmDOI{10.1145/1122445.1122456}

\acmConference[BIOKDD '20]{BIOKDD 20: International Workshop on Data Mining for Bioinformatics}{August 24, 2020}{San Diego, CA}
\acmBooktitle{BIOKDD '20: International Workshop on Data Mining for Bioinformatics, August 24, 2020, San Diego, CA}
\acmPrice{15.00}
\acmISBN{978-1-4503-XXXX-X/18/06}

\begin{document}

\title[SODA: Detecting Covid-19 in Chest X-rays with Semi-supervised Open Set Domain Adaptation]{SODA: Detecting Covid-19 in Chest X-rays with Semi-supervised Open Set Domain Adaptation}

\author{Jieli Zhou}
\authornotemark[1]
\email{zhoujieli777@hotmail.com}
\affiliation{%
  \institution{Carnegie Mellon University}
  \state{PA}
  \country{USA}
}

\author{Baoyu Jing}
\authornote{Both authors contributed equally to this research.}
\email{baoyuj2@illinois.edu}
\affiliation{%
  \institution{University of Illinois at Urbana-Champaign}
  \state{IL}
  \country{USA}
}

\author{Zeya Wang}
\email{zw17.rice@gmail.com}
\affiliation{%
  \institution{Petuum Inc.}
  \state{PA}
  \country{USA}
}

%%%%%%%%%%%%%%%%%%%%
%     Abstract
%%%%%%%%%%%%%%%%%%%%
\begin{abstract}
Due to the shortage of COVID-19 viral testing kits and the long waiting time, radiology imaging is used to complement the screening process and triage patients into different risk levels. Deep learning based methods have taken an active role in automatically detecting COVID-19 disease in chest x-ray images, as witnessed in many recent works in early 2020. Most of these works first train a 
Convolutional Neural Network (CNN) on an existing large-scale chest x-ray image dataset and then fine-tune it with a COVID-19 dataset at a much smaller scale. However, direct transfer across datasets from different domains may lead to poor performance for CNN due to two issues, the large domain shift present in the biomedical imaging datasets and the extremely small scale of the COVID-19 chest x-ray dataset. In an attempt to address these two important issues,
% (domain shift and small training data) and fully exploit the available large-scale well-labeled chest x-ray image datasets,
we formulate the problem of COVID-19 chest x-ray image classification in a semi-supervised open set domain adaptation setting and propose a novel domain adaptation method, \underline{S}emi-supervised \underline{O}pen set \underline{D}omain \underline{A}dversarial network (SODA). SODA is able to align the data distributions across different domains in a general domain space and also in a common subspace of source and target data. In our experiments, SODA achieves a leading classification performance compared with recent state-of-the-art models in separating COVID-19 with common pneumonia. We also present initial results showing that SODA can
produce better pathology localizations in the chest x-rays.
\end{abstract}




\maketitle

%%%%%%%%%%%%%%%%%%%%
%   Introduction
%%%%%%%%%%%%%%%%%%%%
\section{Introduction}
Since the Coronavirus disease 2019 (COVID-19) was first declared as a Public Emergency of International Concern (PHEIC) on January 30, 2020\footnote{\url{https://www.statnews.com/2020/01/30/who-declares-coronavirus-outbreak-a-global-health-emergency/}}, it has quickly evolved from a local outbreak in Wuhan, China to a global pandemic, taking away hundreds of thousands of lives and causing dire economic loss worldwide. In the US, the total COVID-19 cases grew from just one confirmed on Jan 21, 2020 to over 1 million on April 28, 2020 in a span of 3 months. Despite drastic actions like shelter-in-place and contact tracing, the total cases in US kept increasing at an alarming daily rate of 20,000 - 30,000 throughout April, 2020.  
A key challenge for preventing and controlling COVID-19 right now is the ability to quickly, widely and effectively test for the disease, since testing is usually the first step in a series of actions to break the chains of transmission and curb the spread of the disease.

COVID-19 is caused by the severe acute respiratory syndrome coronavirus 2 (SARS-CoV-2) \footnote{\url{https://www.who.int/emergencies/diseases/novel-coronavirus-2019/technical-guidance/naming-the-coronavirus-disease-(covid-2019)-and-the-virus-that-causes-it}}
By far, the most reliable diagnosis is through Reverse Transcription Polymerase Chain Reaction (RT-PCR) \footnote{\url{https://spectrum.ieee.org/the-human-os/biomedical/diagnostics/how-do-coronavirus-tests-work}} in which a sample is taken from the back of throat or nose of the patients and tested for viral RNA. 
While taking samples from the patients, aerosol pathogens could be released and would put the healthcare workers at risk. 
Furthermore, once the sample is collected, the testing process usually takes several hours and recent study reports that the sensitivity of RT-PCR is around 60-70\% \cite{ai2020correlation}, which suggests that many people tested negative for the virus may actually carry it thus could infect more people without knowing it. On the other hand, the sensitivity of chest radiology imaging for COVID-19 was much higher at 97\% as reported by \cite{ai2020correlation, fang2020sensitivity}. 

Due to the shortage of viral testing kits, the long period of waiting for results, and low sensitivity rate of RT-PCR, radiology imaging has been used as a complementary screening process to assist the diagnosis of COVID-19. Unlike RT-PCR, imaging is readily available in most healthcare facilities around the world, and the whole process can be done rapidly. Additionally, one of the key step in reducing mortality rate is early patient triage, since the health level of patients in severe conditions could quickly deteriorate while waiting for the viral testing result. Radiology imaging can provide more detailed information about the patients, e.g. pathology location, lesion size, the severity of lung involvement \cite{zhang2020clinically}. These insights can help doctors to timely triage patients into different risk levels, bring patients in severe conditions to ICU earlier, and saving more lives.

In recent years, with the rapid advancement in deep learning and computer vision, many breakthroughs have been developed in using Artificial Intelligence (AI) for medical imaging analysis, especially disease detection  \cite{wang2017chestx,irvin2019chexpert, wang2018tienet} and report generation \cite{jing2017automatic, li2018hybrid, jing2019show, biswal2020clinical}, and some AI models achieve expert radiologist-level performance \cite{lakhani2017deep}. 
Right now, with most healthcare workers busy at front lines saving lives, the scalability advantage of AI-based medical imaging systems stand out more than ever. Some AI-based chest imaging systems have already been deployed in hospitals to quickly inform healthcare workers to take actions accordingly\footnote{\url{https://spectrum.ieee.org/the-human-os/biomedical/imaging/hospitals-deploy-ai-tools-detect-covid19-chest-scans}}. 

Annotated datasets are required for training AI-based methods, and a small chest x-ray dataset with COVID-19 is collected recently: COVID-ChestXray \cite{cohen2020covid}.
In the last few weeks, several works \cite{wang2020covidnet, li2020artificial, apostolopoulos2020covid, minaee2020deep} apply Convolutional Neural Networks (CNN) and transfer learning to detect COVID-19 cases from chest x-ray images. 
They first train a CNN on a large dataset like Chexpert \cite{irvin2019chexpert} and ChestXray14 \cite{wang2017chestx}, and then fine-tune the model on the small COVID-19 dataset.
By far, due to the lack of large-scale open COVID-19 chest x-ray imaging datasets, most works only used a very small amount of positive COVID-19 imaging samples \cite{cohen2020covid}.
While the reported performance metrics like accuracy and AUC-ROC are high, it is likely that these models overfit on this small dataset and may not achieve the reported performance on a different and larger COVID-19 x-ray dataset.
Besides, these methods suffer a lot from label domain shift: these newly trained models lose the ability to detect common thoracic diseases like ``Effusion'' and ``Nodule'' since these labels do not appear in the new dataset.
Moreover, they also ignored the visual domain shift between the two datasets. 
On the one hand, the large-scale datasets like ChestXray14 \cite{wang2017chestx} and Chexpert \cite{irvin2019chexpert} are collected from top U.S. health institutes like National Institutes of Health (NIH) clinical center and Stanford University, which are well-annotated and carefully processed.
On the other hand, COVID-ChestXray \cite{cohen2020covid} is collected from a very diverse set of hospitals around the world and they are of very different qualities and follow different standards, such as the viewpoints, aspect ratios and lighting, etc. 
In addition, COVID-ChestXray contains not only chest x-ray images but also CT scan images. 

In order to fully exploit the limited but valuable annotated COVID-19 chest x-ray images and the large-scale chest x-ray image dataset at hand, as well as to prevent the above-mentioned drawbacks of those fine-tuning based methods, we define the problem of learning a x-ray classifier for COVID-19 from the perspective of open set domain adaptation (Definition \ref{def:uoda}) \cite{panareda2017open}.
Different from traditional unsupervised domain adaptation which requires the label set of both source and target domain to be the same, the open set domain adaptation allows different domains to have different label sets.
This is more suitable for our problem because COVID-19 is a new disease which is not included in the ChestXray14 or Chexpert dataset.
However, since our task is to train a new classifier for COVID-19 dataset, we have to use some annotated samples.
Therefore, we further propose to view the problem as a Semi-Supervised Open Set Domain Adaptation problem (Definition \ref{def:soda}).

Under the given problem setting, we propose a novel \underline{S}emi-supervised \underline{O}pen set \underline{D}omain \underline{A}dversarial network (SODA) comprised of four major components: a feature extractor $G_f$, a multi-label classifier $G_y$, domain discriminators $D_g$ and $D_c$, as well as common label recognizer $R$.
SODA learns the domain-invariant features by a two-level alignment, namely, domain level and common label level.
The general domain discriminator $D_g$ is responsible for guiding the feature extractor $G_f$ to extract domain-invariant features.
However, it has been argued that the general domain discriminator $D_g$ might lead to false alignment and even negative transfer \cite{pei2018multi, wang2019adversarial}.
For example, it is possible that the feature extractor $G_f$ maps images with ``Pneumonia'' in the target domain and images with ``Cardiomegaly'' in the source domain into similar positions, which might result in the miss-classification of $G_y$.
In order to solve this problem, we propose a novel common label discriminator $D_c$ to guide the model to align images with common labels across domains.
For labeled images, $D_c$ only activates when the input image is associated with a common label. 
For unlabeled images, we propose a common label recognizer $R$ to predict their probabilities of having a common label.

The main contributions of the paper are summarized as follows:
\begin{itemize}
    \item To the best of our knowledge, we are the first to tackle the problem of COVID-19 chest x-ray image classification from the perspective of domain adaptation.
    \item We formulate the problem in a novel semi-supervised open set domain adaptation setting.
    \item We propose a novel two-level alignment model: \underline{S}emi-supervised \underline{O}pen set \underline{D}omain \underline{A}dversarial network (SODA).
    \item We present a comprehensive evaluation to demonstrate the effectiveness of the proposed SODA.
\end{itemize}

%%%%%%%%%%%%%%%%%%%%
%    Preliminary
%%%%%%%%%%%%%%%%%%%%
\section{Preliminary}
\subsection{Problem Definition}

\begin{definition}\label{def:uoda}
\textit{Unsupervised Open Set Domain Adaptation}

Let x be the input chest x-ray image, y be the ground-truth disease label. We define $\mathcal{D}^{s}=\{(\mathbf{x}_n^{s}, y^{s}_n)\}_{n=1}^{N^s}$ as a source domain with $N^s$ labeled samples, and $\mathcal{D}^{t}=\{(\mathbf{x}_n^{t})\}_{n=1}^{N^t}$ as a target domain with $N^t$ unlabeled samples, where the underlying label set $\mathcal{L}^{t}$ of the target domain might be different from the label set $\mathcal{L}^{s}$ of the source domain.  
Define $\mathcal{L}^c = \mathcal{L}^s\cap\mathcal{L}^t$ as the \textit{set of common labels} shared across different domains, 
$\bar{\mathcal{L}}^s=\mathcal{L}^s\textbackslash \mathcal{L}^c$ and $\bar{\mathcal{L}}^t=\mathcal{L}^t\textbackslash \mathcal{L}^c$ be \textit{sets of domain-specific labels} which only appear in the source and the target domain respectively.

The task of \textit{Unsupervised Open Set Domain Adaptation} is to build a model which could accurately assign common labels in $\mathcal{L}^c$ to samples $\mathbf{x}^t_n$ in the target domain, as well as distinguish those $\mathbf{x}^t_n$ belonging to $\bar{\mathcal{L}}^t$.
\end{definition}

\begin{definition}\label{def:soda}
\textit{Semi-supervised Open Set Domain Adaptation} 

Given a source domain $\mathcal{D}^{s}=\{(\mathbf{x}_n^{s}, y^{s}_n)\}_{n=1}^{N^s}$ with $N^s$ labeled samples, and a target domain $\mathcal{D}^{t}\cup\mathcal{D}^{t'}$ consisting of $\mathcal{D}^{t}=\{(\mathbf{x}_n^{t})\}_{n=1}^{N^t}$ with $N^t$ unlabeled samples and $\mathcal{D}^{t'}=\{(\mathbf{x}_n^{t'}, y_n^{t'})\}_{n=1}^{N^{t'}}$ with $N^{t'}$ labeled samples.

The task of \textit{Semi-supervised Open Set Domain Adaptation} is to build a model to assign labels from $\mathcal{L}^t$ to unlabeled samples in $\mathcal{D}^{t}$.

\end{definition}

\subsection{Notations}
We summarize the symbols used in the paper and their descriptions in Table \ref{tab:notations}.

\begin{table}[h]
    \centering
    \caption{Notations}
    \begin{tabular}{l|l}
    \hline
    Symbols & Description \\
    \hline
    \hline
    $\mathcal{D}^{s}$ & set of labeled samples in the source domain \\
    $\mathcal{D}^{t}$ & set of unlabeled samples in the target domain \\
    $\mathcal{D}^{t'}$ & set of labeled samples in the target domain \\
    $\mathcal{L}^s$ & set of labels for the source domain \\
    $\mathcal{L}^t$ & set of labels for the target domain \\
    $\mathcal{L}^c$ & set of common labels across domains \\
    $\bar{\mathcal{L}}^s$ & set of domain-specific labels in the source domain\\
    $\bar{\mathcal{L}}^t$ & set of domain-specific labels in the target domain\\
    $\mathcal{L}$ & set of all labels from all domains\\
    $N^s$ & number of labeled samples in the source domain \\
    $N^t$ & number of unlabeled samples in the target domain \\
    $N^{t'}$ & number of labeled samples in the target domain \\
    \hline
    $G_f$ & feature extractor \\
    $G_y$ & multi-label classifier for $\mathcal{L}$\\
    $G_{y_l}$ & binary classifier for label $l$ (part of $G_y$)\\
    $R$ & common label recognizer\\
    $D_{c}$ & domain discriminator for common labels $\mathcal{L}^c$\\
    $D_{g}$ & general domain discriminator\\
    $L_{G_y}$ & loss of multi-label classification over the entire dataset\\
    $L_R$ & loss of $R$ over the entire dataset\\
    $L_{D_g}$ & loss of $D_g$ over the entire dataset\\
    $L_{D_c}$ & loss of $D_c$ over the entire dataset\\
    $\lambda$ & the coefficient of losses\\
    \hline
    $\mathbf{x}$ & input image\\
    $\mathbf{h}$ & hidden features\\
    $y$ & ground-truth label\\
    $\hat{y}$ & predicted probability\\
    $\hat{d}$ & predicted probability that $\mathbf{x}$ belongs to source domain\\
    $\hat{r}$ & predicted probability that $\mathbf{x}$ has common labels \\
    \hline
    \end{tabular}
    \label{tab:notations}
\end{table}

%%%%%%%%%%%%%%%%%%%%
%    Methodology
%%%%%%%%%%%%%%%%%%%%

\section{Methodology}
\subsection{Overview}

\begin{figure*}[t]
    \centering
    \includegraphics[width=.95\textwidth]{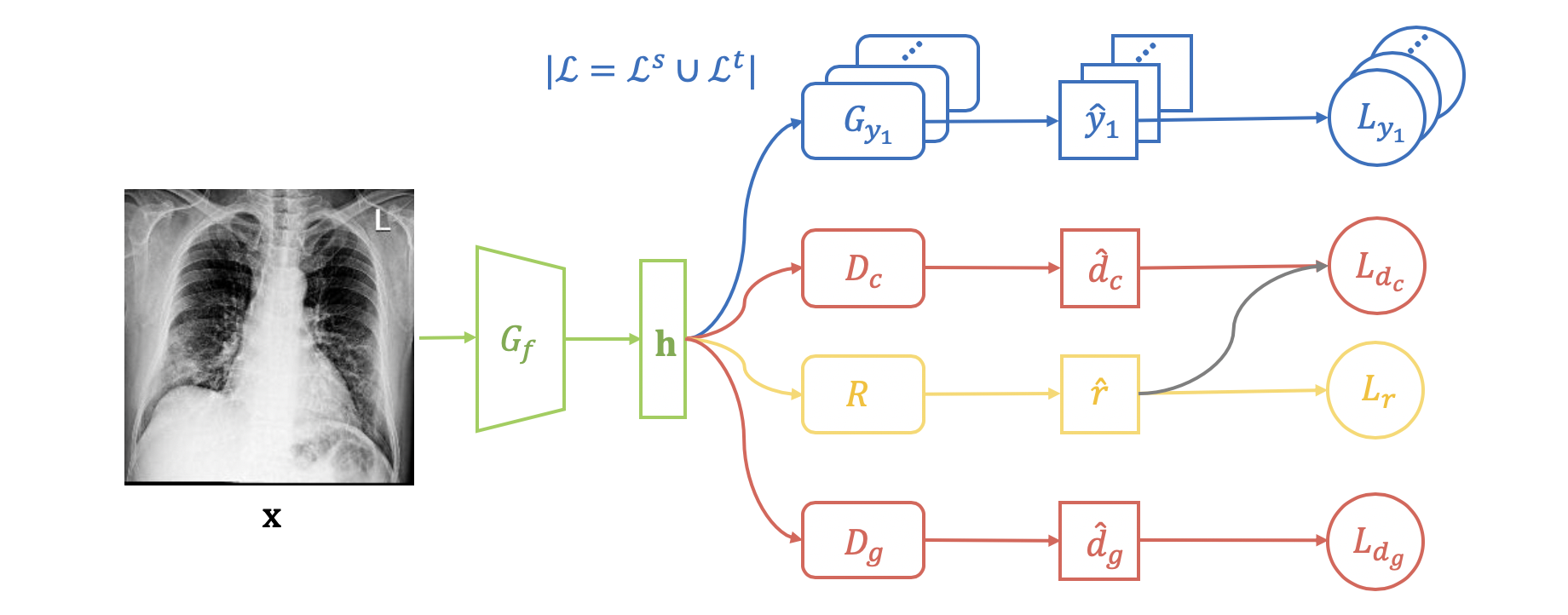}
    \caption{Architecture overview of the proposed SODA model. Given an input image $\mathbf{x}$, the feature extractor $G_f$ will extract its hidden features $\mathbf{h}$ (green part), which will be fed into a multi-label classifier $G_y$ (blue part), a common label recognizer $R$ (yellow part) and a domain discriminator $D$ (red part) to predict the probability $\hat{y}$ of disease labels, the probability $\hat{r}$ that $\mathbf{x}$ is associated with a common label  and the probability $\hat{d}$ that $\mathbf{x}$ belongs to the source domain. 
    $L_y$, $L_r$ and $L_d$ denote the losses of image classification, common label classification and domain classification.
    $D_g$ is the general domain discriminator, and $D_c$ is the domain discriminator for images associated with a common label.
    $G_{y_1}$ denotes the image classifier for the first label in the label set of the entire dataset $\mathcal{L}=\mathcal{L}^s\cup\mathcal{L}^t$.
    Note that the gradients from $L_{d_c}$ and $L_{d_g}$ are not allowed to pass through $\hat{r}$ (grey arrows).
    }
    \label{fig:model}
\end{figure*}

An overview of the proposed \underline{S}emi-supervised \underline{O}pen Set \underline{D}omain \underline{A}dversarial network (SODA) is shown in Fig. \ref{fig:model}.
% The model is comprised of four major components: an image feature extractor $G_f$ (green part), a multi-label image classifier $G_y$ (blue part), a common label recognizer $R$ (yellow part), and a domain discriminator $D$ (red part).
% The domain discriminator $D$ can be further divided into two parts: one ($D_g$) for general domain adaptation, and one ($D_{l}$) for aligning distributions of images associated with common labels $l\in\mathcal{L}^c$ across domains.
Given an input image $\mathbf{x}$, it will be first fed into a feature extractor $G_f$, which is a Convolutional Neural Network (CNN), to obtain its hidden feature $\mathbf{h}$ (green part).
The binary classifier $G_{y_l}$ (part of the multi-label classifier $G_y$) takes $\mathbf{h}$ as input, and will predict the probability $\hat{y}_l$ for the label $l\in\mathcal{L}$ (blue part).

We propose a novel two-level alignment strategy for extracting the domain invariant features across the source and target domain. 
On the one hand, we perform \textit{domain alignment} (Section \ref{sec:align_domain}), which leverages a general domain discriminator $D_g$ to minimize the domain-level feature discrepancy. 
On the other hand, we emphasize the \textit{alignment of common labels} $\mathcal{L}^c$ (Section \ref{sec:align_common}) by introducing another domain discriminator $D_c$ for images associated with common labels. 
For labeled images in $\mathcal{D}^s$ and $\mathcal{D}^{t'}$, we compute loss for $D_c$ and conduct back-propagation during training only if the input image $\mathbf{x}$ is associated with a common label $l\in\mathcal{L}^c$.
As for unlabeled data in $\mathcal{D}^{t}$, we propose a common label recognizer $R$ to predict the probability $\hat{r}$ that an image $\mathbf{x}$ has a common label, and use $\hat{r}$ as a weight in the losses of $D_c$ and $D_g$.

\subsection{Domain Alignment}\label{sec:align_domain}
Domain adversarial training \cite{ganin2016domain} is the most popular method for helping feature extractor $G_f$ learn domain-invariant features such that the model trained on the source domain can be easily applied to the target domain.
The objective function of the domain discriminator $D_g$ can be written as:
\begin{equation}\label{eq:loss_D_g}
\begin{split}
    L_{D_g} =& -\mathbb{E}_{(\mathbf{x}^s\in\mathcal{D}^s)}[\log\hat{d}_g] \\
    &-\mathbb{E}_{(\mathbf{x}^t\in\mathcal{D}^t\cup\mathcal{D}^{t'})}[\log(1 -\hat{{d}}_g)]
\end{split}
\end{equation}
where $\hat{{d}}_g$ denotes the predicted probability that the input image belongs to the source domain.

In SODA, we use a Multi-Layer Perceptron (MLP) as the general domain discriminator $D_g$.

\subsection{Common Label Alignment} \label{sec:align_common}
In the field of adversarial domain adaptation, most of the existing methods only leverage a general domain discriminator $D_g$ to minimize the discrepancy between the source and target domain.
Such a practice ignores the label structure across domains, which will result in false alignment and even negative transfer \cite{pei2018multi,wang2019adversarial}.
If we only use a general domain discriminator $D_g$ in the open set domain adaptation setting (Definition \ref{def:uoda} and Definition \ref{def:soda}), it is possible that the feature extractor $G_f$ will map the target domain images with a common label $l\in\mathcal{L}^c$, ``Pneumonia'', and the source domain images with a specific label $l\in\bar{\mathcal{L}}^s$, e.g., ``Cardiomegaly'', to similar positions in the hidden space, which might lead to the classifier miss-classifying a ``Pneumonia'' image in the target domain as ``Cardiomegaly''.

To address the problem of the miss-matching between the common and specific label sets, we propose a domain discriminator $D_c$ to distinguish the domains for the images with a common label.
For the labeled data from the source domain $\mathcal{D}^s$ and the target domain $\mathcal{D}^t$, we know whether an image $\mathbf{x}$ has a common label or not, and we only calculate the loss $L_{d_c}$ for $D_c$ on the samples with common labels:
\begin{equation}\label{eq:loss_D_c_label}
\begin{split}
    L_{D_c}^{label} =& -\mathbb{E}_{(\mathbf{x}^s\in\mathcal{D}^s, y^s\in\mathcal{L}^c)}[\log\hat{d}_c] \\
    &-\mathbb{E}_{(\mathbf{x}^t\in\mathcal{D}^{t'}, y^{t'}\in\mathcal{L}^c)}[\log(1 -\hat{{d}}_c)]
\end{split}
\end{equation}
where $\hat{d}_c$ denotes the predicted probability that the input images is associated with a common label.

However, a large number of images in the target domain are unlabeled, and thus extra effort is required for determining whether an unlabeled image is associated with a common label.
To address this problem, we propose a novel common label recognizer $R$ to predict the probability $\hat{r}$ whether an unlabeled image has at least one common label.
The probability $\hat{r}$ will be used as a weight in the loss function of $D_c$\footnote{Note that gradients stop at $\hat{r}$ in the training period.}:

\begin{equation}\label{eq:loss_D_c_un}
    L_{D_c}^{un} = -\mathbb{E}_{(\mathbf{x}^t\in\mathcal{D}^{t}, y^{t}\in\mathcal{L}^c)}[\hat{r}\log(1 -\hat{{d}}_c)]
\end{equation}

In addition, we also use $\hat{r}$ to re-weigh unlabeled samples in $D_g$ (Equation \ref{eq:loss_D_g}) to further emphasize the alignment of common labels:

\begin{equation}\label{eq:loss_D_g_2}
\begin{split}
    L_{D_g} =& -\mathbb{E}_{(\mathbf{x}^s\in\mathcal{D}^s)}[\log\hat{d}_g] \\
    & -\mathbb{E}_{(\mathbf{x}^t\in\mathcal{D}^{t'})}[\log(1 -\hat{{d}}_g)] \\
    & -\mathbb{E}_{(\mathbf{x}^t\in\mathcal{D}^t)}[\hat{r}\log(1 -\hat{{d}}_g)] 
\end{split}
\end{equation}
Finally, the recognizer $R$ is trained on the labeled set $\mathcal{D}^s\cup\mathcal{L}^{t'}$ via cross-entropy loss:
\begin{equation}\label{eq:loss_R}
\begin{split}
    L_{R} =& -\mathbb{E}_{(\mathbf{x}\in\mathcal{D}^s\cup\mathcal{D}^{t'}, y\in\mathcal{L}^c)}[\log\hat{r}] \\
    &-\mathbb{E}_{(\mathbf{x}\in\mathcal{D}^s\cup\mathcal{D}^{t'}, y\notin\mathcal{L}^c)}[\log(1 -\hat{{r}})]
\end{split}
\end{equation}

\subsection{Overall Objective Function}
The overall objective function of SODA can be written as a min-max game between classifiers $G_y$, $R$ and discriminators $D_g$, $D_c$:

\begin{equation}
    \min_{G_y, R}\max_{D_g, D_c} L_{G_y} + \lambda_R L_{R} - \lambda_{D_g}L_{D_g} - \lambda_{D_c}^{label}L_{D_c}^{label} - \lambda_{D_c}^{un}L_{D_c}^{un}
\end{equation}

\noindent where $L_R$, $L_{D_g}$, $L_{D_c}^{label}$ and $L_{D_c}^{un}$ are respectively defined in Equation \ref{eq:loss_R}, \ref{eq:loss_D_g_2}, \ref{eq:loss_D_c_label} and \ref{eq:loss_D_c_un}; $L_{G_y}$ denotes the cross-entropy loss for multi-label classification; $\lambda$ denotes the coefficient for each loss function.
%to trade-off between transferability and discriminability.

%%%%%%%%%%%%%%%%%%%%
%    Experiments
%%%%%%%%%%%%%%%%%%%%

% \newpage
\section{Experiments}
\subsection{Experiment Setup}
\subsubsection{Datasets}
\paragraph{Source Domain}
We use ChestXray-14 \cite{wang2017chestx} as the source domain dataset. This dataset is comprised of 112,120 anonymized chest x-ray images from the National Institutes of Health (NIH) clinical center.
The dataset contains 14 common thoracic disease labels: ``Atelectasis'', ``Consolidation'', ``Infiltration'', ``Pneumothorax'', ``Edema'', ``Emphysema'', ``Fibrosis'', ``Effusion'', ``Pneumonia'', ``Pleural thickening'', ``Cardiomegaly'', ``Nodule'', ``Mass'' and ``Hernia''.

\paragraph{Target Domain}
The newly collected COVID-ChestXray \cite{cohen2020covid} is adopted as the target domain dataset. This dataset contains images collected from various public sources and different hospitals around the world. This dataset (by the time of this writing) contains 328 chest x-ray images in which 253 are labeled positive as the new disease ``COVID-19'', whereas 61 are labeled as other well-studied ``Pneumonia''.

\subsubsection{Evaluation Metrics}
We evaluate our model from four different perspectives.
First, to test the classification performance, following the semi-supervised protocol, we randomly split the 328 x-ray images in COVID-ChestXray into 40\% labeled set, and 60\% unlabeled set. 
We run each model 3 times and report the average AUC-ROC score.
Second, we compute the Proxy-$\mathcal{A}$ Distance (PAD) \cite{ben2007analysis} to evaluate models' ability for minimizing the feature discrepancy across domains.
Thirdly, we use t-SNE to visualize the feature distributions of the target domain.
Finally, we also qualitatively evaluate the models by visualizing their saliency maps.

\subsubsection{Baseline Methods}
We compare the proposed SODA with two types of baselines methods: fine-tuning based transfer learning models and domain adaptation models.
For fine-tuning based models, we select the two most popular CNN models DenseNet121 \cite{huang2017densely} and ResNet50 \cite{he2016deep} as our baselines. 
These models are first trained on the ChestXray-14 dataset and then fine-tuned on the COVID-ChestXray dataset.
As for the domain adaptation models, we compare our model with two classic models, Domain Adversarial Neural Networks (DANN) \cite{ganin2016domain} and Partial Adversarial Domain Adaptation (PADA) \cite{cao2018partial}. 
Note that DANN and PADA were designed for unsupervised domain adaptation, and we implement a semi-supervised version of them.

\subsubsection{Implementation Details}
We use DenseNet121 \cite{huang2017densely}, which is pretrained on the ChestXray-14 dataset \cite{wang2017chestx}, as the feature extractor $G_f$ for SODA.
The multi-label classifier $G_y$ is a one layer neural network and its activation is the sigmoid function.
We use the same architecture for $D_g$, $D_c$ and $R$: a MLP containing two hidden layers with ReLU \cite{nair2010rectified} activation and an output layer.
The hidden dimension for all of the modules: $G_y$, $D_g$, $D_c$ and $R$ is 1024.
For fair comparison, we use the same setting of $G_f$, $G_y$ and $D_g$ for DANN \cite{ganin2016domain} and PADA \cite{cao2018partial}.
All of the models are trained by Adam optimizer \cite{kingma2014adam}, and the learning rate is $10^{-4}$.

\subsection{Classification Results}
To investigate the effects of domain adaptation and demonstrate the performance improvement of the proposed SODA, we present the average AUC-ROC scores for all models in Table \ref{tab:results}. Comparing the results for ResNet50 and DenseNet121, we observe that deeper and more complex models achieve better classification performance. For the effects of domain adaptation, it is obvious that the domain adaptation methods (DANN, PADA, and SODA) outperform those fine-tuning based transfer learning methods (ResNet50 and DenseNet121).
% The average score for non-domain adaptation methods is 0.8173 for COVID-19 and 0.8378 for Pneumonia, while the average score for domain adaptation method is 0.8871 for COVID-19 and 0.9027 for Pneumonia. 
% We observe an 8.5\%  improvement (absolute improvement of 0.0698) in detecting COVID-19 and a 7.7\% improvement (absolute improvement of 0.0649) in detecting Pneumonia. 
Furthermore, the proposed SODA achieves higher AUC scores on both COVID-19 and Pneumonia than DANN and PADA, demonstrating the effectiveness of the proposed two-level alignment.

\begin{table}[h]
    \centering
    \caption{Target Domain Average AUC-ROC Score}
    \begin{tabular}{l|l|l}
    \hline
    Model & COVID-19 & Pneumonia \\
    \hline
    ResNet50 \cite{he2016deep} & 0.8143 & 0.8342 \\
    DenseNet121 \cite{huang2017densely} & 0.8202 & 0.8414  \\
    
    \hline
    DANN \cite{ganin2016domain} & 0.8785 & 0.8961 \\
    PADA \cite{cao2018partial} & 0.8822 & 0.9038\\
    \hline
    SODA & \textbf{0.9006} & \textbf{0.9082}\\
    \hline
    \end{tabular}
    \label{tab:results}
\end{table}

\subsection{Proxy $\mathcal{A}$-Distance}
Proxy $\mathcal{A}$-Distance \cite{ben2007analysis} has been widely used in domain adaptation for measuring the feature distribution discrepancy between the source and target domains.
PAD is defined by 
\begin{equation}
d_{\mathcal{A}}=2(1-2\min(\epsilon))    
\end{equation}
where $\epsilon$ is the domain classification error (e.g. mean absolute error) of a classifier (e.g. linear SVM \cite{cortes1995support}).

\begin{figure}[h!]
    \centering
    \includegraphics[width=0.4\textwidth]{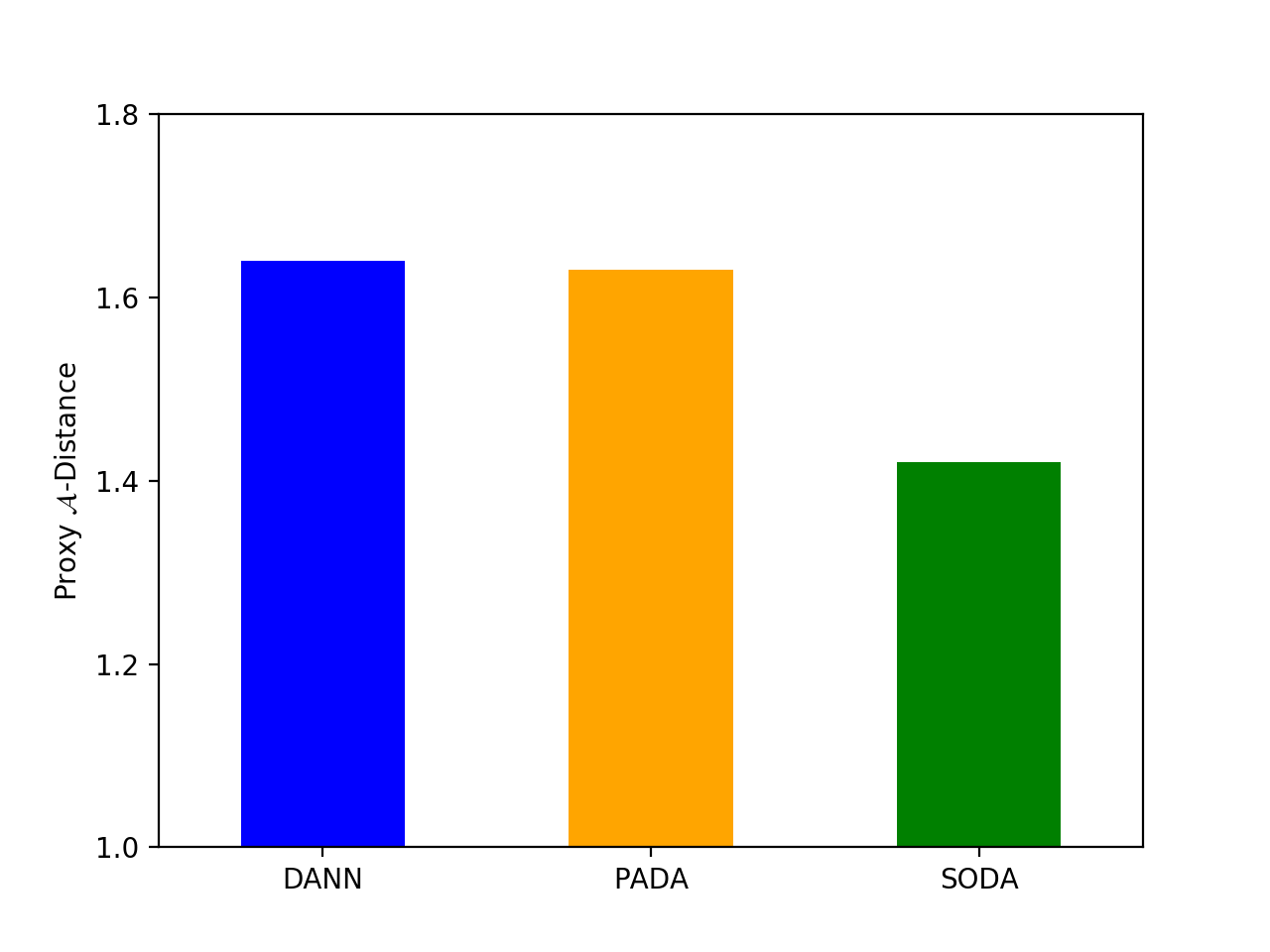}
    \caption{Proxy $\mathcal{A}$-Distance}
    \label{fig:pad}
\end{figure}

Following \cite{ganin2016domain}, we train SVM models with different $C$ and use the minimum error to calculate PAD.
In general, a lower $d_\mathcal{A}$ means a better ability for extracting domain invariant features.
As shown in Fig. \ref{fig:pad}, SODA has a lower PAD compared with the baseline methods, which indicates the effectiveness of the proposed two-level alignment strategy.

\begin{figure*}[t!]
\centering
\subfloat[DANN]
{\includegraphics[width=0.32\linewidth]{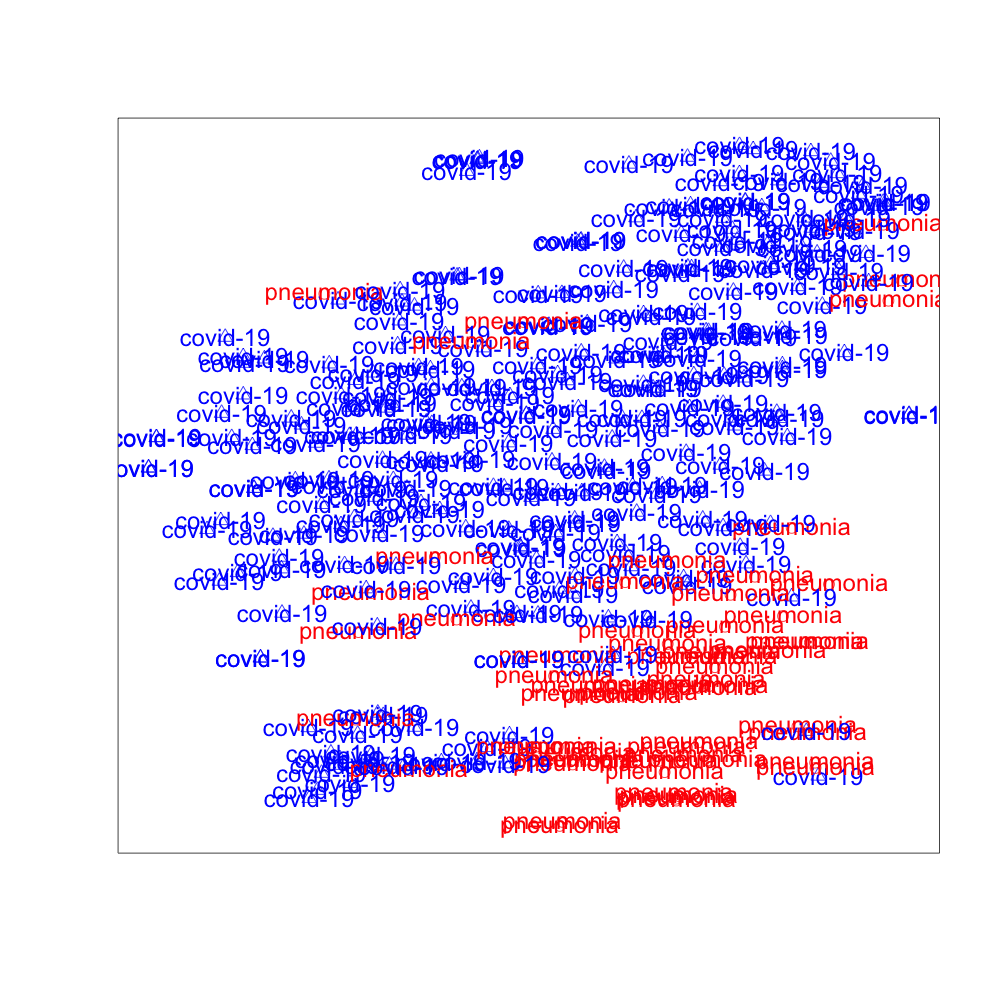}}\quad
\subfloat[PADA]
{\includegraphics[width=0.32\linewidth]{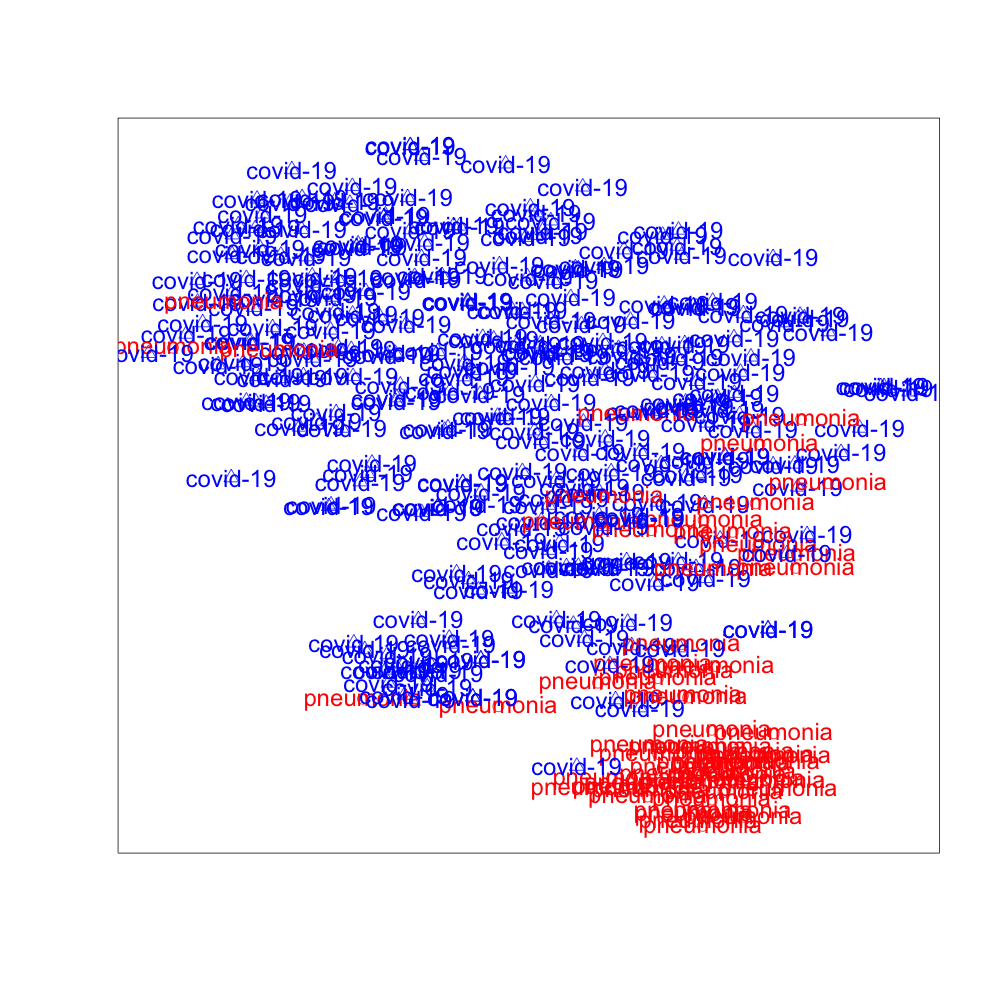}}\quad
\subfloat[SODA]
{\includegraphics[width=0.32\linewidth]{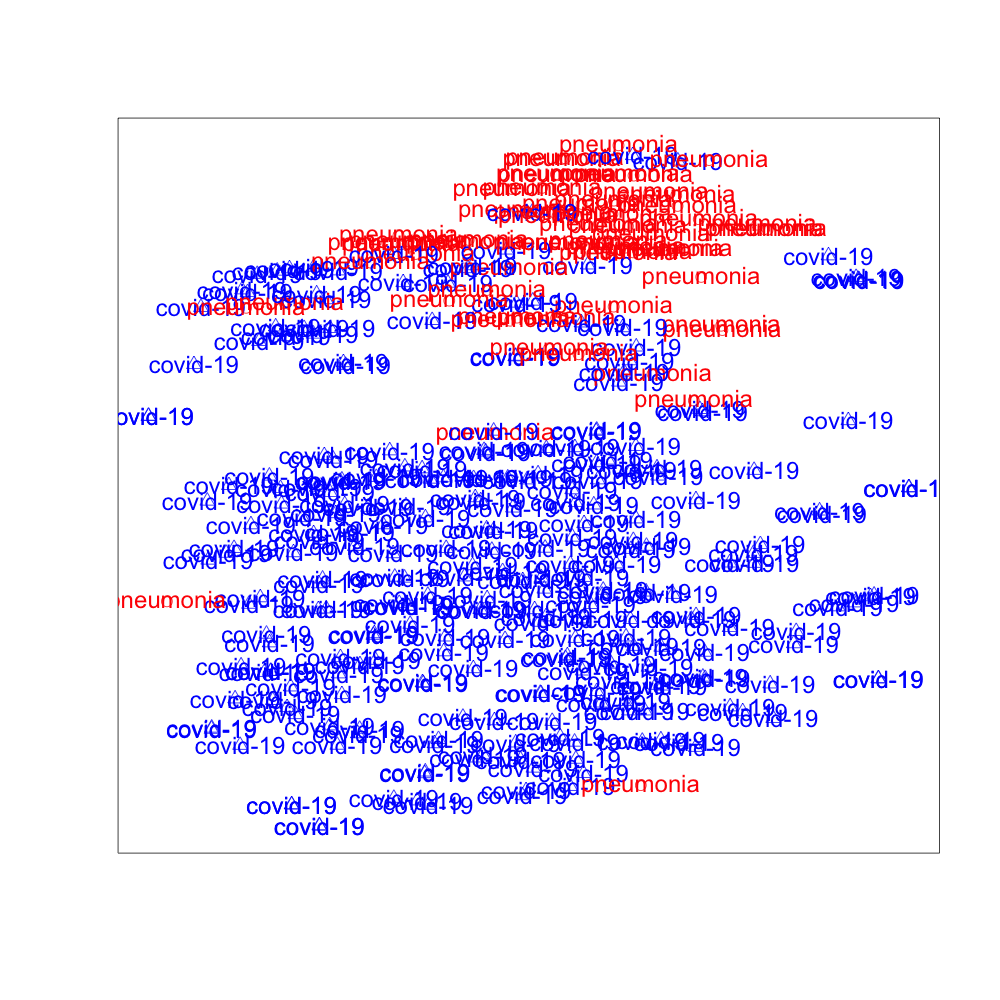}}\quad
\caption{t-SNE visualization for DANN, PADA and SODA on the target domain.}\label{fig:tsne}
\end{figure*}

\begin{figure*}[!htb]
    \centering
    \includegraphics[width=.9\textwidth]{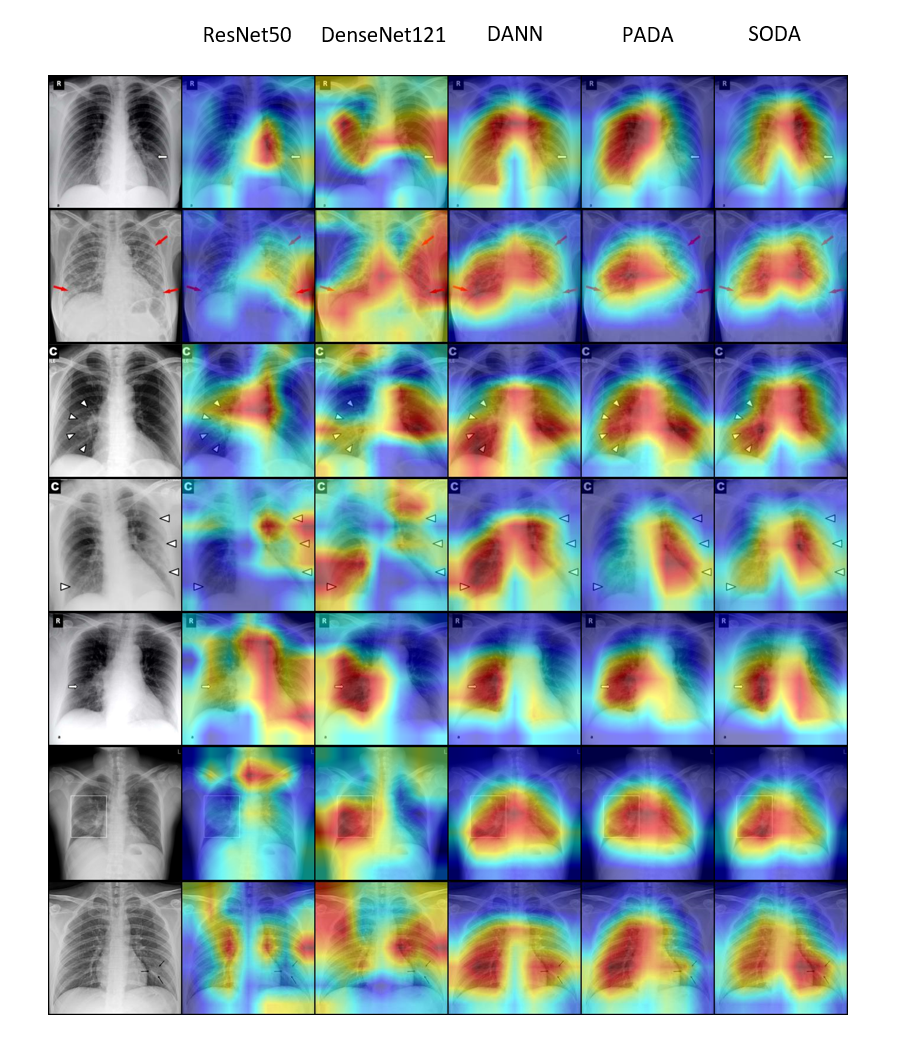}
    \caption{Grad-CAM\cite{selvaraju2017grad} visualization for ResNet50, DenseNet121, DANN, PADA and SODA. From left to right, the first column is the chest x-ray images from COVID-ChestXray dataset, the second and third columns are visualization of the weights on last layers from ResNet50 and DenseNet121, the fourth and fifth columns are visualizations of weights from the last layers of the domain adaptation models DANN and PADA. The last column is the visualization of our model. SODA has the best pathology localization among all models. 
    }
    \label{fig:gradcam}
\end{figure*}

\subsection{Feature Visualization}
We use t-SNE to project the high dimensional hidden features $\mathbf{h}$ extracted by DANN, PADA, and SODA to low dimensional space.
The 2-dimensional visualization of the features in the target domain is presented in Fig. \ref{fig:tsne}, where the red data points are image features of ``Pneumonia'' and the blue data points are image features of ``COVID-19''.
It can be observed from Fig. \ref{fig:tsne} that SODA performs the best for separating ``COVID-19'' from ``Pneumonia'', which demonstrates the effectiveness of the proposed common label recognizer $R$ as well as the domain discriminator for common labels $D_c$. 
% Additionally, there are some images labeled ``COVID-19'', which might suggest some cases in this dataset are mild cases share image features with those of Pneumonia. 

\subsection{Grad-CAM}
Grad-CAM \cite{selvaraju2017grad} is used to visualize the features extracted from all compared models. Fig. \ref{fig:gradcam} shows the Grad-CAM results on seven different COVID-19 positive chest x-rays. These seven images have annotations (small arrows and box) indicating the pathology locations. We observe that ResNet50 and DenseNet121 can focus wrongly on irrelevant locations like the dark corners and edges. In contrast, domain adaptation models have better localization in general, and our SODA model gives more focused and accurate pathological locations than other models compared. In addition, we consult a professional radiologist with over 15 years of clinical experience from Wuxi People's Hospital and received positive feedback on the pathological locations as indicated by the Grad-CAM of SODA. In the future, we plan to do a more rigorous evaluation study with more inputs from radiologists. We believe the features extracted from SODA can assist radiologists to pinpoint the suspect COVID-19 pathological locations faster and more accurately.

%%%%%%%%%%%%%%%%%%%%
%    Related Work
%%%%%%%%%%%%%%%%%%%%
\section{Related Work}

\subsection{Domain Adaptation}
Domain adaptation is an important application of transfer learning that attempts to generalize the models from source domains to unseen target domains \cite{ganin2015unsupervised,ganin2016domain,tzeng2017adversarial,tzeng2014deep,you2019universal, jing2018cross, wang2020coarse}. 
Deep domain adaptation approaches are usually implemented through discrepancy minimization~\cite{tzeng2014deep} or adversarial training~\cite{ganin2015unsupervised,ganin2016domain,tzeng2017adversarial}. 
Adversarial training, inspired by the success of generative adversarial modeling~\cite{goodfellow2014generative}, has been widely applied for promoting the learning of transfer features in image classification. 
It takes advantage of a domain discriminator to classify whether an image is from the source or target domains. 
% The adversarial training is usually done to ensure the distributions of visual features learned over different domains to be similar in two ways, one is through a gradient reversal layer (GRL) for simultaneously training the feature extractors and domain discriminators ~\cite{ganin2015unsupervised}, and the other is through an iterative updating scheme between feature extractors and domain discriminators  ~\cite{tzeng2017adversarial}. 
On top of these methods, a couple of works have been presented for exploring the high-level structure in the label space, which aim at further improving the domain adaptation performance for multi-class image classification ~\cite{wang2019adversarial} or fundamentally solving the application problem when the label sets from source domains and target domains are different ~\cite{you2019universal}. 
In order to meet the latter target, more and more researchers have started to study the open set domain adaptation problem, in which case the target domain has images that do not come from the classes in the source domain ~\cite{you2019universal, panareda2017open}. Universal domain adaptation is the latest method that is proposed through using an adversarial domain discriminator and a non-adversarial domain discriminator to successfully solve this problem.~\cite{you2019universal}. 
Although domain adaptation has been well explored, its application in medical imaging analysis, such as domain adaptation for chest x-ray images, is still under-explored. 

\subsection{Semi-supervised Learning}
Semi-supervised learning is a very important task for image classification, which can make use of both labeled and unlabeled data at the same time \cite{saito2019semi}. 
Recently it has been used to solve image classification problems on a very large (1 billion) set of unlabelled images \cite{yalniz2019billion}. 
% Supervised approaches typically require labeled examples in the target domain and unsupervised approaches do not use any labeled images in the target domain, while semi-supervised methods take into account both the labeled and unlabeled target images ~\cite{saito2019semi}.
In spite of many progresses that have been made with unsupervised domain adaptation methods, the domain adaptation with semi-supervised learning has not yet been fully explored. 

\subsection{Chest X-Ray Image Analysis}
There has been substantial progress in constructing publicly available databases for chest x-ray images as well as a related line of works to identify lung diseases using these images. 
The largest public datasets of chest x-ray images are Chexpert \cite{irvin2019chexpert} and ChestXray14 \cite{wang2017chestx}, which respectively include more than 200,000 and 100,000 chest x-ray images collected by Stanford University and National Institute of Healthcare. 
The creation of these datasets have also motivated and promoted the multi-label chest x-ray classification for helping the screening and diagnosis of various lung diseases. 
The problems of disease detection \cite{wang2017chestx,irvin2019chexpert, wang2018tienet} and report generation using chest x-rays \cite{jing2017automatic, li2018hybrid, jing2019show, biswal2020clinical} are fully investigated and have achieved much-improved results upon recently. 
However, there have been very few attempts for studying the domain adaptation problems with the multi-label image classification problem using chest x-rays.

%%%%%%%%%%%%%%%%%%%%
%    Conclusion 
%%%%%%%%%%%%%%%%%%%%
\section{Conclusion}
In this paper, in order to assist and complement the screening and diagnosing of COVID-19, we formulate the problem of COVID-19 chest x-ray image classification in a semi-supervised open set domain adaptation framework. Accordingly, we propose a novel deep domain adversarial neural network, \underline{S}emi-supervised \underline{O}pen set \underline{D}omain \underline{A}dversarial network (SODA), which is able to align the data distributions across different domains at both domain level and common label level. Through evaluations of the classification accuracy, we show that SODA achieves better AUC-ROC scores than the recent state-of-the-art models. We further demonstrate that the features extracted by SODA is more tightly related to the lung pathology locations, and get initial positive feedback from an experienced radiologist. In practice, SODA can be generalized to any semi-supervised open set domain adaptation settings where there are a large well-annotated dataset and a small newly available dataset. In conclusion, SODA can serve as a pilot study in using techniques and methods from domain adaptation to radiology imaging classification problems.

\bibliographystyle{ACM-Reference-Format}
\bibliography{sample-base}

\end{document}